\documentclass[12pt]{iopart}

\usepackage{graphicx}
\usepackage{color, colortbl}
\usepackage[minnames=1,maxnames=4,doi=false,isbn=false,style=nature,eprint=false,backend=biber,natbib=true,hyperref=true]{biblatex}
\definecolor{Gray}{gray}{0.9}			

\DeclareUnicodeCharacter{2212}{-}
\begin{document}

\title[Author guidelines for IOP Publishing journals in  \LaTeXe]{Perspectives on testing fundamental physics with highly charged ions in Penning traps}

\author{K. Blaum, S. Eliseev \& S. Sturm}

\address{Max-Planck-Institute for Nuclear Physics, Saupfercheckweg 1, 69117 Heidelberg, Germany}
\ead{klaus.blaum@mpi-hd.mpg.de}
\vspace{10pt}
\begin{indented}
\item[]February 2020
\end{indented}

\begin{abstract}
In Penning traps electromagnetic forces are used to confine charged particles under well-controlled conditions for virtually unlimited time. Sensitive detection methods have been developed to allow observation of single stored ions. Various cooling methods can be employed to reduce the energy of the trapped particle to nearly at rest. In this review we summarize how highly charged ions offer unique possibilities for precision measurements in Penning traps. Precision atomic and nuclear masses as well as magnetic moments of bound electrons allow among others to determine fundamental constants like the mass of the electron or to perform stringent tests of fundamental interactions like bound-state quantum electrodynamics. Recent results and future perspectives in high-precision Penning-trap spectroscopy with highly charged ions will be discussed.  
\end{abstract}


%
%
%
%
%

\section{Introduction}

A single highly charged ion (HCI) in a Penning trap is a unique system to investigate various facets of fundamental physics. Two parameters of a HCI are of particular interest in this investigation: the mass of a HCI and the magnetic moments of its atomic electrons. The Penning trap is a tool that enables a high-precision determination of the HCI parameters via measuring under well-controlled conditions the characteristic frequencies of the HCI motion confined to a very small volume in a Penning trap.\\
The mass of a HCI is an integral characteristic that reflects the masses of its constituents and their interactions. One gets access to the energetics of various nuclear reactions via measuring the masses of the involved nuclides and particles and hence gets a handle on their properties. One prominent example are experiments on the determination of the neutrino mass and the search for keV sterile neutrinos. These experiments \cite{ECHo,HOLMeS,SterileNeutrino} can greatly benefit from an independent determination of the $Q$-values of certain beta-processes (the mass differences of the parent and daughter nuclides of the processes). Another example is a model-independent test of the equality of the speed of light and the limiting speed of massive matter which is considered a direct test of special relativity \cite{Gree1991}. This test is based on the investigation of the energy balance of the non-resonant capture of a cold neutron by certain nuclides \cite{Jent2018}. The crucial parameter to be known precisely in this investigation is a mass difference of the nuclides participating in the reaction.\\
HCIs are key players in the development of new-generation ion clocks \cite{KozlovRMP2018}. They can possess meta-stable electron configurations that exhibit just a very weak sensitivity to external electromagnetic perturbations and have configuration energies of a few ten to hundred eV. These two factors can facilitate a creation of a very stable ion clock for, e.g., probing the time variation of fundamental constants \cite{safronova2018search}. Thus, a search for such meta-stable electron configurations is of great importance and can be accomplished by a  direct measurement of the mass differences of a nuclide of interest in the same charge state but in different electron configurations \cite{Rima2020}.\\
Highly charged heavy ions with just a few electrons in the atomic shell are probably at the moment the most suitable objects to test the validity of Quantum Electrodynamics (QED) in a regime of strong electromagnetic fields. For instance, the innermost electron in heavy nuclides like lead or uranium is exposed to a Coulomb field of the nucleus of approximately 10$^{16}$ V/cm, the value which is hardly achievable in laboratories by other means. In such a strong Coulomb field QED effects manifest themselves in the binding energies and magnetic moments ($g$-factors) of the considered electrons on a level measurable by Penning-trap techniques.\\



\section{Basics of Penning traps and ion cooling}

\subsection{Ion motion in a Penning trap}

\begin{figure*}[h!]
\includegraphics[width=1\textwidth]{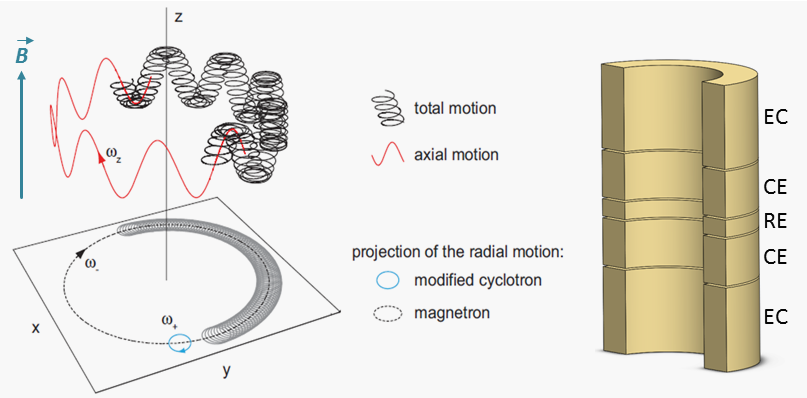}
\centering
\caption{(left) The ion motions in a Penning trap: (1) axial motion - a periodic oscillation with frequency $\omega_z=2\pi\nu_z$ in the $z$-direction in the trap electrostatic potential, (2) cyclotron and magnetron motions with frequencies $\omega_+=2\pi\nu_+$ and $\omega_-=2\pi\nu_-$, respectively, in the $xy$-plane perpendicular to the magnetic field lines. (right) Cylindrical Penning trap - a set of usually five ring electrodes of equal diameter that allows a creation of sufficiently quadratic electrostatic potential in the volume of ion confinement.}
\label{trap_motion}
\end{figure*}

The Penning trap  is a superposition of two static fields - a strong uniform magnetic field $B$ and a weak quadrupolar electrostatic field created by two electrodes shaped as infinite hyperboloids of revolution. The magnetic field confines the motion of a charged particle (ion) with mass $m$ and charge $q$ to a plane perpendicular to the magnetic field lines (radial plane) whereas the quadrupolar electrostatic  potential well constrains the ion motion along the magnetic field lines. In a Penning trap an ion performs three independent motions - cyclotron, magnetron and axial motions - with frequencies $\nu_{+}, \nu_{-}$ and $\nu_{z}$, respectively  (see Figure~\ref{trap_motion}). 
The cyclotron motion is a circular motion of an ion around the magnetic field lines. The axial motion is a harmonic oscillation of an ion along the magnetic field lines in the quadrupolar electrostatic potential well. The magnetron motion is a slow circular drift of an ion in the radial plane caused by the cross product of the magnetic and electrostatic fields. The frequency hierarchy is $\nu_{c}>\nu_{+}\gg\nu_{z}\gg\nu_{-}$ for typical trap parameters and ion masses.\\ 
The recipe for determination of the ion mass is given by the invariance theorem \cite{Brown1982}, which is applicable not only to an ideal Penning trap, but also to existing Penning-trap systems with small imperfections in the alignment of the electric and magnetic fields and with a non-vanishing ellipticity of the trap electrodes
\begin{equation}
\frac{1}{2\pi} \frac{q}{m} B = \nu_{c}=\sqrt{\nu^{2}_{+}+\nu^{2}_{-}+\nu^{2}_{z}},
\label{eq:Invariance}
\end{equation}
where $\nu_{c}$ is the  frequency which the ion cyclotron motion would have in the pure magnetic field $B$. \\
High-precision Penning-trap mass measurements on short-lived nuclides employ a simpler relation of the cyclotron frequency to the trap eigenfrequencies 
\begin{equation}
\nu_{c}=\nu_{+}+\nu_{-}.
\label{eq:SideBand}
\end{equation}
It allows a determination of the mass ratio of two ions with close masses with a relative uncertainty down to at least few parts in $10^{10}$ for all existing on-line Penning-trap facilities \cite{Gab2009}.\\

Besides Penning-trap mass spectrometers there is a class of ultra-precise Penning-trap experiments devoted to the determination of the $g$-factor of charged particles.
A charged particle with mass $m_s$, charge $q_s$ and a non-zero spin performs in a magnetic field $B$ a spin precession with Larmor frequency 
\begin{equation}
\nu_L = \frac{1}{2\pi} \frac{g}{2} \frac{q_s}{m_s} B.
\label{eq:Larmor}
\end{equation}
Thus, in order to determine the $g$-factor, it is necessary to measure the Larmor frequency and the magnetic field strength $B$. For example, in the case of a $g$-factor of an electron bound in a highly charged ion with known mass $m$ and charge $q$, the magnetic field strength is determined by measuring the cyclotron frequency $\nu_c$ of the ion. Such a combined measurement yields for the $g$-factor of a bound electron the following expression:
\begin{equation}
g = 2\frac{\nu_L}{\nu_c} \frac{m_s}{m} \frac{q}{q_s}.
\label{eq:g_factor}
\end{equation}

A real Penning trap always differs from the ideal case. This leads to the fact that the measured eigenfrequency values in general exhibit certain non-vanishing offsets from the ideal Penning-trap values. There is quite a number of effects in the real Penning trap that can cause such frequency offsets. The most important ones are a misalignment of the trap and magnetic field axes, an ellipticity of the trap, a non-harmonicity of the electrostatic trap potential, and an inhomogeneity of the magnetic field. Also the image-charge effect and the relativistic dependence of the ion cyclotron frequency on its kinetic energy becomes non-negligible in the case of ultra-precise frequency-ratio measurements on highly charged ions.\\ 

\subsection{Cooling of the ion motions in a Penning trap}

A prerequisite for high-precision measurements of the ion cyclotron frequency is a confinement of the ion motions in the Penning trap to a small volume. This is achieved by various "ion cooling" techniques. The term "ion cooling" means in this context a reduction of the ion motion amplitudes. There are four distinctively different ion cooling techniques that are employed in Penning traps: (1) mass-selective buffer gas cooling \cite{Sav1991}, (2) electron cooling \cite{Rol1989,Ber2004}, (3) resistive cooling \cite{Win1975} and (4) laser cooling \cite{Lar1986}. \\
The mass-selective buffer gas cooling technique is a method of choice in online Penning-trap experiments \cite{SHIPTRAP,TRIGATRAP,JYFLTRAP,ISOLTRAP,LEBIT,TITAN,APT}.  Short-lived nuclides of interest are synthesized in nuclear reactions in the form of high-energy highly charged ions, are slowed down in a gas-filled stopping chamber to thermal energy reducing their charge state to 1+, 2+ and rarely 3+, are converted in a gas-filled linear Paul trap from a continuous beam of low charged ions into ion bunches and finally are delivered to a Penning-trap mass spectrometer, which is usually a tandem of two Penning traps: cooler trap and measurement trap. The mass-selective buffer gas cooling technique is applied in the cooler trap, which is filled with helium gas at a typical pressure of a few 10$^{-5}$ mbar and a room temperature of about 300 K. The cyclotron and axial ion motions can be considered stable oscillators. Thus, in the cooler trap they come into thermal equilibrium with helium atoms. The amplitude of the magnetron motion being metastable increases due to collisions with helium atoms. The "cooling" of the magnetron motion is achieved by transferring its action to the cyclotron motion by means of a $\pi$-pulse at the sideband frequency $\nu_+$+$\nu_-$. As a result, the "cooled" low charged ions in the cooler trap have cyclotron and magnetron amplitudes of just about 0.1 mm, whereas the amplitude of the axial motion does not exceed 1 mm. The virtue of this cooling technique is that it allows for fast cooling (sub-100 ms) of trap motions of low charged ions at room temperature to sub-mm amplitudes. The largest disadvantage of the technique is that it is not applicable to highly charged ions.\\
One Penning-trap facility, namely TITAN \cite{TITAN}, stands out from the group of online Penning-trap mass spectrometers in that it employs highly charged ions to measure their masses. Since a "cooling" of highly charged ions in neutral gas is not possible, an electron cooling technique is proposed instead \cite{Rol1989,Ber2004}. The TITAN's cooler trap contains a plasma of electrons that are injected into the cooler trap from an electron gun. Highly charged ions trapped in the cooler trap are sympathetically cooled by the co-trapped electron plasma to a common energy without bonding with the electrons.\\
In Penning traps that aim at a measurement of the ion cyclotron and Larmor frequencies with a relative uncertainty well below $10^{-10}$ a resistive cooling technique is a working-horse method of decreasing the ion trap-motion amplitudes to a $\mu m$ scale \cite{Win1975}. The resistive cooling is realized by connecting to one of the trap electrodes a superconducting NbTi resonator with a very high quality factor. The inductance of the resonator and the capacitance of the trap form a resonant circuit which resonance frequency is usually tuned to that of the ion's axial or cyclotron motion. The resonator has a certain impact on the ion motion. From one side, the voltage induced on the electrodes by the ion motion creates across the trap an electric field that tends to decrease the amplitude of the ion motion. From the other side, the thermal noise of the resonator tends to excite the ion motion. As a result, the ion motion finally comes into a thermal equilibrium with the resonator. Since the resonator and the trap are usually kept at a temperature of liquid helium, the amplitude of the thermalized ion motion does not exceed a few $\mu m$.\\
For some applications temperatures below the 4K environment are required. This is the case in the very strongly inhomogeneous magnetic bottle fields required for the detection of the spin state  via the continuous Stern-Gerlach effect in $g$-factor experiments (see chapter \ref{eq:g_factor}). Here, especially the residual cyclotron energy causes sizeable shifts of the axial frequency and can consequentially cause an unacceptable frequency instability if not kept constant carefully. As the heating rate due to residual noise on the electrodes scales approximately linearly with the total cyclotron energy \cite{bohman2018sympathetic} it is advantageous or even mandatory to cool the motion to very low temperatures. Similarly, for high-precision laser spectroscopy, the linear Doppler shift resulting from the axial motion is the dominant source of line broadening if the ion is not cooled into or close to the Lamb-Dicke regime where individual motional sidebands can be resolved.\\
Doppler laser cooling of trapped alkali-like ions such as Be$^+$ and Mg$^+$ can reach temperatures of about 0.5~mK or even lower. Since the ion is trapped, a single laser beam (for an almost closed transition) is sufficient to cool e.g. the axial motion. Radiofrequency coupling of the motional sidebands can then cool the other modes as well. Generally however, HCI do not feature fast optical transitions suitable for direct laser cooling. Consequentially, these ions of interest have to be cooled sympathetically via interaction with directly cooled ions. This interaction can take place either via Coulomb interaction for co-trapped ions, or via the induced image charges in shared electrodes. The latter technique has the important advantage that the motional frequencies of the HCI in that case are not significantly modified. This however comes at the cost of a much weaker coupling strength and consequentially slower cooling rate, which can only partially be counteracted by using not a single but a cloud of typically some hundred Be$^+$ ions. Currently, advanced techniques for the efficient coupling of the two species are in development at several experiments. Once successful, this will pave the way to ultra-high precision experiments on various HCI using laser and microwave spectroscopy as well as mass spectrometry.

\subsection{Measurement of the free cyclotron and Larmor frequencies}

There is a considerable difference in methods employed at different classes of Penning traps for a measurement of the free cyclotron frequency. Each class is characterized with certain distinct features, which are listed in Table~\ref{tab:Features}.
\begin{table} [h!]
\centering
\caption{A list of features which distinctly characterize each class of high-precision Penning traps.}
 \label{tab:Features}
\begin{tabular*}{1\textwidth}{@{\extracolsep{\fill}}cc}
\hline
\hline
on-line facilities for experiments & off-line setups for experiments\\
 on short-lived nuclides &  on long-lived or stable nuclides \\
\hline
\rowcolor{Gray}
operated at room temperature & operated at 4.2 K \\
typically low charged ions used & highly charged ions used \\
\rowcolor{Gray}
ions with half-lives as short as a few ms & very long-lived or stable ions \\
very low on-line production & off-line production of ions\\
rates of nuclides &  in large quantity \\
\rowcolor{Gray}
uncertainty in mass determination &  uncertainty in mass-ratio determination \\
\rowcolor{Gray}
is a few 100 eV/c$^2$ up to a few keV/c$^2$ &  is below a few eV/c$^2$ \\
frequency-measurement technique: &  frequency-measurement technique: \\
ToF-ICR, PI-ICR; destructive &  FT-ICR; non-destructive \\
\hline
\hline
\end{tabular*}
\end{table}

At online facilities two cyclotron-frequency techniques are in use: the Time-of-Flight Ion-Cyclotron-Resonance (ToF-ICR) \cite{ToF1980,ToF1995,ToF1992,ToF2007_1,ToF2007_2,ToF2007_3} and Phase-Image Ion-Cyclotron-Resonance (PI-ICR) \cite{PIICR_1,PIICR_2} techniques. The ToF-ICR has been the method of choice until now due to its relative simplicity and sensitivity to a single ion in the trap at room temperature. Since this method is destructive, at least a few hundred ions are required to reach a sufficiently low uncertainty in mass determination.  The recently developed novel Phase-Image Ion-Cyclotron-Resonance (PI-ICR) technique is approximately by a factor of 25 faster than the ToF-ICR method and offers by a factor of 40 higher resolving power. It has a perspective to replace the ToF-ICR technique in mass measurements on very short-lived nuclides and low lying nuclear isomeric states \cite{Manea2020}.\\
In cryogenic off-line setups a set of frequency-measurement techniques united under a common name "the Fast-Fourier Ion-Cyclotron-Resonance (FT-ICR) technique" is employed \cite{FTICR_1,FTICR_2,FTICR_3,FTICR_4,FTICR_5}. It is based on a measurement of the frequency of a current (image current) induced by the ion trap motion in a superconducting resonant tank which is attached to one of the trap electrodes. This technique allows frequency measurements on a single ion with an extraordinary low uncertainty. The price to pay for this is a high complexity of the system, which requires a cooling of the trap as well as the associated frequency-measurement electronics to the temperature of liquid helium.\\ 
In experiments on the determination of the $g$-factor one has to measure besides the free cyclotron frequency also the Larmor frequency of the spin precession. Since the spin precession is not associated with a charge oscillation, the techniques based on the image-current frequency measurement are not applicable. Instead, a technique that make use of the continuous Stern-Gerlach method is employed \cite{dehmelt1986continuous,Stur2019}. It is based on the detection of the spin flip between two spin states of a charged particle of interest, which is placed in a Penning trap with strong quadratic magnetic inhomogeneity.\\      

\section{Production and properties of highly charged ions}
While the Penning trap supports precision spectroscopy with arbitrary charged atomic and molecular ions, especially for fundamental physics applications a high charge state $Z$ can be a decisive advantage. This can be either simply due to the larger charge-to-mass ratio, which leads to a larger cyclotron frequency and correspondingly improved measurement precision, or due to the special atomic structure of such ions. When removing most of the electrons of an atom, the resulting HCI has a drastically simpler electron shell, which allows for high-precision atomic physics calculations. Moreover, the remaining electrons are then bound much stronger to the nucleus, where they experience the strongest electromagnetic fields accessible today in the laboratory in stable systems - up to $10^{16}$~V/cm electric and 20,000~T magnetic fields. Accordingly, HCI are close to ideal candidates for tests of the fundamental laws of physics, the Standard Model, in extreme conditions. Furthermore, the presence of the strong field decreases the susceptibility to external influences such as stray electric and magnetic fields, again enabling even higher precision.
The strong fields also govern the scaling behaviour of the energy levels in the HCI. The expectation value of the electron radius moves closer to the nucleus with $1/Z$, and correspondingly the electronic transitions of the gross structure scale roughly proportional to $Z^2$, which quickly shifts them into the x-ray and even gamma regime. On the other hand, the fine- and hyperfine structure transitions, which are typically in the low microwave range of frequencies for atoms, scale by $Z^4$ and $Z^3$, respectively, which makes them accessible to precision laser spectroscopy studies in a suitable range of $Z$. Only the Zeeman splittings, which originate from an externally applied magnetic field, remain in the millimeter regime (for typical 2-7 Tesla magnetic fields) throughout the complete range of $Z$. This makes these transitions and the associated $g$-factor especially versatile for fundamental physics tests.
On the other hand, the high $Z$ of course also increases the ionisation potential and correspondingly imposes significant difficulty for the production and storage of HCI. In the extreme cases of hydrogenlike Pb$^{81+}$ or Bi$^{82+}$ the energy required to remove the 1s electrons is higher than 100 keV. Such energies require specialized large-scale installations as there exist only a few world-wide. One possibility is to accelerate singly charged ions to GeV energies and then to strip the other electrons in a metallic foil. This procedure, as it is implemented at the experimental storage ring (ESR) at the GSI Helmholtzzentrum für Schwerionenforschung in Darmstadt, Germany \cite{franzke1987heavy}, results in HCI produced in flight. To make these ions accessible for Penning-trap experiments they have to be cooled and decelerated almost to rest. This elaborate procedure is currently developed at the HITRAP facility at GSI \cite{kluge2008hitrap}. 
The alternative to the ion-accelerator based method is to use a high-energy electron beam impinging on the ions. As the energy transfer in electron-electron collisions is drastically more efficient than in the ion-electron case, here, energies in the 200~keV range suffice. This principle is used in electron-beam ion traps (EBITs). Here, the ions are confined in a strong magnetic field by means of cylindrical electrostatic electrodes as well as the space-charge of the electron beam which extends axially along the magnetic field lines. The trapped ions are then successively charge-bred to higher charge states, limited only by the available electron energy and recombination with rest-gas atoms from the imperfect vacuum. 
Technically, however, it is challenging to supply strong electron beams in the order of Amp\`eres with the required $>$100~keV energies. Currently, only very few machines world-wide operate in this regime - one of those is the Heidelberg Super-EBIT which is currently upgraded to enable the highest energies. The produced ion cloud consists of a distribution of charge-states at a typically quite high temperature owing to the collisions with the electron beam. However, unlike in the accelerator case, the energy distribution is typically only in the several eV range, allowing a direct capturing in a cryogenic Penning trap without preceding cooling steps, albeit with finite capturing efficiency. At the ALPHATRAP setup, ions are extracted from the HD-EBIT, selected by q/m and injected into the cryogenic trap with typically a few percent efficiency, absolutely enough considering the experiment requires only a single ion. In hermetically sealed cryogenic traps the vacuum can be so good that recombinations are basically eliminated, allowing for storage times in the order of months even for HCI. In this case it becomes feasible to keep a number of the ions of interest in a dedicated "reservoir" trap \cite{Stur2019}. This way, if an ion gets lost in the measurement or setup procedure, a new one can be extracted from the reservoir without the need to run the EBIT again, which drastically increases the productivity.

\section{Precision Penning-trap mass spectrometry}
\label{ch:MassSpec}
High-precision atomic and nuclear mass data have broad applications in numerous fields of research including nuclear and nuclear astrophysics, neutrino physics, metrology, tests of fundamental interactions and their symmetries as well as physics beyond the Standard Model. In the following some exciting examples will be discussed in more details.

\subsection{Nuclear masses for test of special relativity}
Special relativity is one of the deepest principles in our understanding of modern physics. Its prominent role has motivated many tests of its foundations and predictions. These tests of special relativity can be divided in several categories according to which assumption they investigate: the isotropy of space, the independence of the speed of light from the velocity of the laboratory, the time dilation effect, and the equality of two distinct “speeds of light“ – the velocity of propagation of an electromagnetic wave in vacuum, c, and the limiting velocity of a massive particle, cm. Only the latter does not assume the existence of a preferred frame of reference and hence can be considered a direct test of special relativity. 
In 1991, G.L. Greene et al. proposed to test this equality, i.e., $c = c_m$ as special relativity predicts, by measuring the wavelength of a photon emitted in a transition where a mass $\Delta m$ is converted into electromagnetic radiation \cite{Gree1991}. From energy conservation it follows that $\Delta mc_m^2 = hc/\lambda$, where $\lambda$ is the photon wavelength and $h$ is the Planck constant. One considers the non-resonant capture of a cold  neutron in the reaction $^AX(n,\gamma)^{A+1}X$: from energy conservation one obtains $[M(^AX) – M(^{A+1}X) + M(n)] c_m^2 = hc/\lambda_{A+1}$, where $\lambda_{A+1}$ is the wave length of a $\gamma$-ray emitted by the nuclide $^{A+1}X$ after a neutron capture by the nuclide $^AX$. The mass difference of two nuclides in the reaction can be measured very precisely with Penning-trap mass spectrometry, whereas the right tool to measure the wavelength is crystal Bragg spectroscopy \cite{Jent2018}.
It took more than ten years until such a test was carried out by combining very accurate measurements of atomic-mass ratios of $^{29}$Si/$^{28}$Si and $^{33}$S/$^{32}$S with the MIT-trap \cite{Corn1989} with a relative uncertainty of about $10^{-11}$, and of $\gamma$-ray wavelengths in the GAMS4 experiment \cite{Kess2001} with fractional accuracy of about $10^{-7}$. The test yielded a result of $1 - c_m/c = -1.4 (4.4)\cdot10^{-7}$, indicating that the velocity of propagation of an electromagnetic wave in vacuum and the limiting velocity of a massive particle are equal to a level of at least $4\cdot10^{-7}$ \cite{Rain2005}. To our knowledge, this is the most precise direct test of the famous equation yet described. 
The sensitivity of about $4\cdot10^{-7}$ on the equality of two velocities achieved in the MIT-trap and GAMS4 experiments is limited by the inability of the GAMS4 Bragg spectrometer to measure the wavelength of a $\gamma$-ray with a fractional uncertainty better than $10^{-7}$. An order of magnitude improved precision in the $\gamma$-ray measurement and thus in the test of special relativity seems to be possible with the upgraded GAMS6 spectrometer \cite{Günt2017}. However, a suitable nuclide must be selected, which should fulfil the following requirements: (1) it must have a relatively long lifetime on the order of at least several weeks to provide precise mass values, (2) it must have a large neutron capture cross section, (3) a mechanical form or chemical compound must be available, which can resist high temperatures inside the reactor, (4) the branching rate of a single cascade must be sufficiently high and (5) the $\gamma$-decay scheme of the daughter nuclide in the reaction must be relatively simple. According to these requirements, the pair $^{35}$Cl/$^{36}$Cl is one of the most promising in case the mass ratio gets measured with the required uncertainty of about $10^{-11}$. Presently, this mass ratio is known with a fractional uncertainty of only about $10^{-9}$ \cite{Audi2017}. We aim to improve the mass uncertainties of $^{35,36}$Cl by two orders of magnitude. The measurement will be carried out with PENTATRAP Penning-trap mass spectrometer \cite{Repp2012} using the laser Tip-EBIT technique for the production of highly charged Cl ions \cite{Schw2019}.

\subsection{Nuclear masses for fifth force search}
\begin{figure*}[h!]
\includegraphics[width=0.5\textwidth]{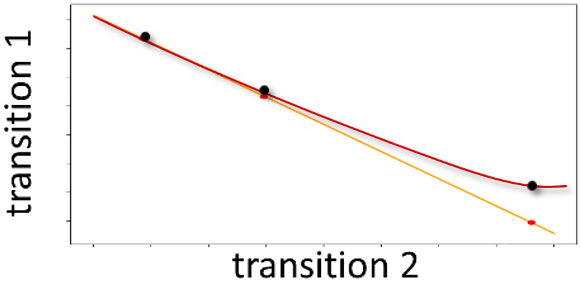}
\centering
\caption{Illustration of linearity (orange) and nonlinearity (red) in the King plot of the isotope shifts in transition 1 versus transition 2.}
\label{fifth_force}
\end{figure*}

Recently, it has been suggested that ultrahigh-precision determinations of isotope shifts based on high-resolution laser spectroscopy can be used to detect interesting new particle candidates \cite{Bere2018}. To that end the linearity of King plot isotope shift comparisons gets tested (see Figure~\ref{fifth_force}) and provides bounds on the existence of new light force mediators which can improve existing bounds even by orders of magnitude \cite{Frug2017}.  

The optical isotope shift between two isotopes $A$ and $A'$ can be written as:
\begin{equation}
\nu_i^{AA'} = K_i\mu_{AA'}+F_i\delta<r^{2}>_{AA'},
\label{eq:IS}
\end{equation}
with 
\begin{equation}
\mu_{AA'}=\frac{1}{m_A}-\frac{1}{m_{A'}},
\label{eq:Massterm}
\end{equation}
where the masses of isotopes $A$ and $A'$ are denoted as $m_A$ and $m_{A'}$, respectively. The first term in Eq.~(\ref{eq:IS}) expresses the mass shift (normal as well as specific mass shift), the second term the leading-order contribution to the field shift, which scales with the difference of mean squares of nuclear charge radii between isotopes $A$ and $A'$. In order not to be limited by the masses of the involved isotopes, high-precision Penning-trap mass measurements on the isotopic chains of interest, e.g. Ca, Sr, and Yb \cite{Solaro2020,Counts2020}, needs to be performed with relative mass uncertainties of $\delta m/m=1\cdot10^{-11}$, as recently demonstrated by PENTATRAP \cite{Rischka2020}.

\subsection{Nuclear masses for neutrino physics studies}

\begin{figure*}[h!]
\includegraphics[width=1\textwidth]{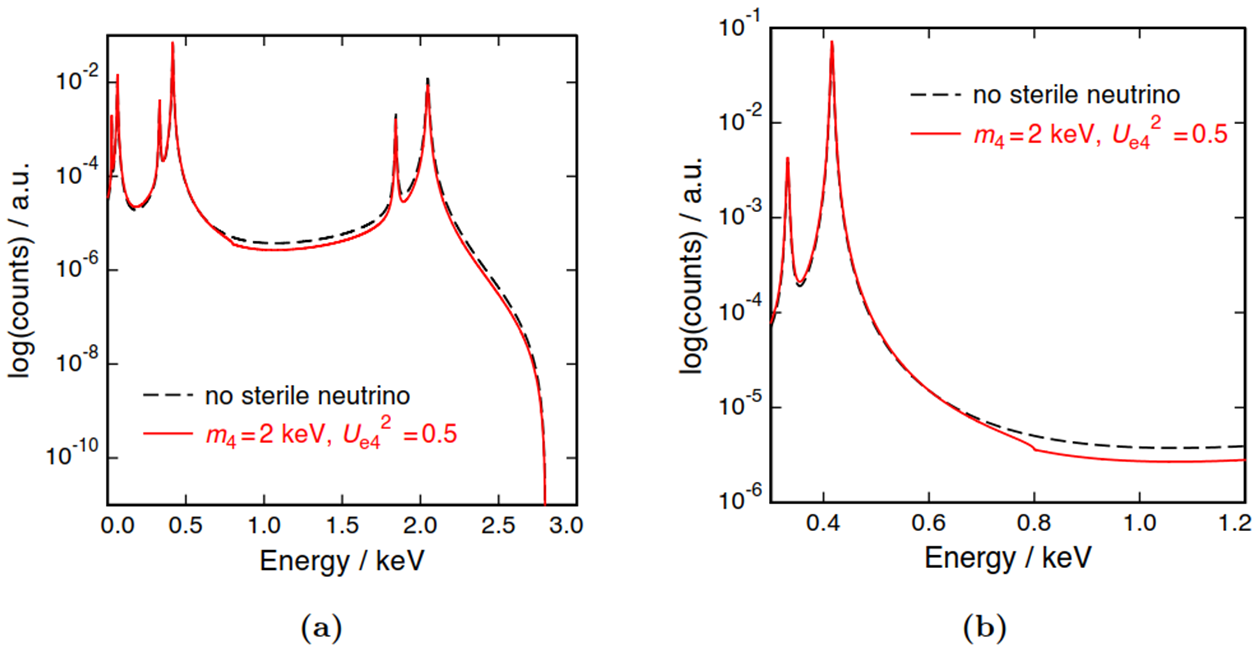}
\centering
\caption{(a) Comparison between the expected calorimetrically measured $^{163}$Ho spectrum in the case of no sterile neutrino (black dashed line) and in the case of a heavy neutrino mass $m_4$ = 2 keV with a mixing of $U^2_{e4}$ = 0.5. (b) A magniﬁcation of (a) in the region of the kink. The plots are taken from \cite{SterileNeutrino}.}
\label{kink}
\end{figure*}

A precise knowledge of the $Q$-values of a variety of nuclear single and double beta processes is essential for addressing such open issues in neutrino physics as (1) the neutrino mass value, (2) the neutrino type and (3) the existence of keV-sterile neutrinos.\\
\indent From several established approaches for the determination of the neutrino mass the investigation of single beta processes ($\beta^-$ decay and electron capture (EC)) is the most convenient and moreover the only model-independent method to determine the mass of the electron neutrino - one of the neutrino flavor states. Several experiments currently struggle to reach a sub eV/c$^2$ uncertainty in the determination of the (anti)neutrino mass from the analysis of the $\beta^-$ decay of tritium \cite{KATRIN,Project8} and the EC in $^{163}$Ho \cite{ECHo,HOLMeS}. The most advanced experiment on the determination of the anti-neutrino mass from the tritium $\beta^-$ decay - the Karlsruhe Tritium Neutrino experiment KATRIN - has recently obtained a value of 1.1 eV/c$^2$ (90\% confidence level) for an upper limit on the absolute mass scale of neutrinos \cite{KATRIN2}. Presently, this is the most stringent limit on the electron neutrino mass ever reached. The KATRIN project can benefit from a very precise direct and independent measurement of the $Q$-value of tritium $\beta^-$ decay. It can help to assess the systematic uncertainty in the neutrino-mass determination more reliably if the $Q$-value be measured with an uncertainty better than a few meV/c$^2$. The FSU-trap has recently advanced towards this by having determined the $Q$-value with an uncertainty of 70 meV via measuring the mass difference of tritium and $^3$He \cite{FSU-trap}. Another suitable beta process for a precise determination of the neutrino mass - the electron capture in $^{163}$Ho - underlies two experiments, ECHo \cite{ECHo} and HOLMES \cite{HOLMeS}. The neutrino mass is determined here from the analysis by means of cryogenic microcalorimetry electron de-excitation processes which follow the electron capture. Two Penning-trap experiments, PENTATRAP \cite{Repp2012} and CHIP-trap \cite{CHIP-trap}, have on their measurement list the determination of the $Q$-value of the EC in $^{163}$Ho with an uncertainty of approximately 1 eV. This will facilitate the determination of the neutrino mass on a sub eV/c$^2$ precision level.\\
\indent It is not clear yet whether the neutrino and antineutrino are different (Dirac) or identical (Majorana) particles. Presently, the most suitable way to determine the neutrino type is the investigation of extremely rare double beta processes. Nuclides which can undergo these processes are in fact virtually stable. They can emit two electrons (2$\beta^-$ decay) or can capture two atomic orbital electrons (2EC) either with or without an emission of two neutrinos. Of interest is only the neutrinoless mode (0$\nu$2$\beta^-$ decay and 0$\nu$2EC), since it  can occur only if neutrinos are Majorana particles. Until now all large-scale experiments have been devoted to the search for 0$\nu$2$\beta^-$ decay \cite{Double_Beta}, since in general, it is by far more probable than 0$\nu$2EC. In these experiments one searches in the energy-sum spectrum of two emitted electrons for a peak at an energy which corresponds to the $Q$-value of the investigated 2$\beta^-$ decay. The $Q$-value needs to be known with an uncertainty of approximately one keV in order (1) to localize the searched peak in the two-electron spectrum and (2) to compute with sufficient precision the nuclear matrix element of the process. All most prominent $2\beta^-$ decay  transitions have been already addressed by various high-precision Penning-trap facilities by determining their $Q$-values  with suﬃcient accuracy \cite{136Xe,76Ge,130Te,100Mo,82Se,116Cd,48Ca,150Nd,96Zr}. 
The other neutrinoless double beta process, 0$\nu$2EC, is in general so improbable that it is considered unobservable. Nevertheless, the probability of certain 0$\nu$2EC transitions can be resonantly enhanced to such extent that it can become comparable to that of $2\beta^-$ decay  transitions. The degree of the resonant enhancement of a certain 0$\nu$2EC transition is solely defined by the difference of the binding energy of two captured electrons and the  $Q$-value of the transition.
Three 0$\nu$2EC transitions between the ground states and 12 0$\nu$2EC transitions from excited states to the ground state  were considered to possess a certain degree of resonant enhancement. Their $Q$-values have been determined with high-precision Penning traps with uncertainties ranging from a few 100 eV down to a few 10 eV \cite{2EC2012}. This extensive experimental $Q$-value measurement campaign has found two at least partially resonantly enhanced 0$\nu$2EC-transitions, $^{152}$Gd$\rightarrow^{152}$Sm \cite{Ga152} and $^{156}$Dy$\rightarrow^{156}$Gd \cite{Dy156}. Their expected half-lives normalized to the effective Majorana neutrino mass value of 1 eV exceed 10$^{26}$ years rendering it impossible to observe the 0$\nu$2EC-decay of these nuclides by means of present experimental technique. \\  
\indent Along with three active neutrinos one predicts the existence of sterile neutrinos - presumably Warm Dark Matter particles - with the mass ranging from 1 to 100 keV \cite{SterileNeutrino}. Although sterile neutrinos do not undergo weak interaction, they mix with active neutrinos with certain mixing angles and thus reveal their existence in beta processes which are accompanied by an emission of neutrinos. One of the most suitable beta processes for searching for keV-sterile neutrinos is electron capture. The ECHo project on the determination of the neutrino mass plans to extend their measurement program to searching for keV-sterile neutrinos by analyzing the cryogenic microcalorimetric atomic de-excitation spectrum of the EC in $^{163}$Ho \cite{SterileNeutrino}. If keV-sterile neutrinos exist, they manifest themselves as a small kink in the spectrum at the energy that corresponds to the difference of the $Q$-value of the process and the mass of the sterile neutrino (see Figure~\ref{kink}).
The EC in $^{163}$Ho is sensitive to sterile neutrinos with mass smaller than 2.5 keV. The employment in the ECHo or similar experiment other EC transitions with larger $Q$-values will allow for a search for keV-sterile neutrinos with masses up to 100 keV. The list of the most suitable EC-transitions for the search for keV-sterile neutrinos includes, for instance, the EC in $^{157}$Tb, $^{163}$Ho, $^{179}$Ta, $^{193}$Pt, $^{205}$Pb \cite{SterileNeutrino}. The precision in the determination of the position of the kink and hence the sterile neutrino mass is directly related to the uncertainty of the determination of the $Q$-value. Thus, an independent and direct determination of the $Q$-values of the above-mentioned EC-transitions is required on a level of at most a few eV-uncertainty and falls into the scope of the PENTATRAP project.\\

\subsection{Determination of electron binding energies and test of QED}

The validity of quantum electrodynamics (QED), so far the best confirmed theory, is persistently questioned in theory and experiments on various atomic systems.
Of particular interest are hydrogen-like ions of heavy nuclides (Pb$^{81+}$, U$^{91+}$). These ions are simple systems of a single electron in a strong Coulomb field (10$^{16}$ V/cm) of the ion nucleus. Thus, these systems give a unique opportunity to test QED in the strongest electromagnetic fields available in experiments.
Unlike light-ion systems QED calculations of hydrogen-like heavy ions have to be performed non-perturbatively in $\alpha Z$ regime. This imposes serious difficulties on these calculations requiring experimental investigations of such systems. One of such investigations is a measurement of the binding energy of the electron in hydrogen-like heavy ions, which allows for the determination of the Lamb shift - the difference between the exact binding energy and the point-nucleus Dirac binding energy.
The Lamb shift in H-like uranium is presently known with an uncertainty of 4.6 eV \cite{LambShift}, thus providing a test of QED in the strong Coulomb field on a 2$\%$ level.
The PENTATRAP experiment plans to substantially reduce the uncertainty in the electron binding energy determination by measuring the mass difference of hydrogen-like and bare heavy nuclides with a sub-eV/c$^2$ uncertainty.\\

\section{Precision magnetic moment measurements of bound electrons}
The extraordinary control over the motion of single trapped ions in Penning traps, combined with novel cooling techniques, allows achieving impressive measurement precision of the magnetic moment of bound electrons in highly charged ions with numerous applications, which will be discussed in the following.

\subsection{Stringent test of bound-state quantum electrodynamics}
The determination of the $g$-factor of the electron bound in hydrogen like ions allows for stringent tests of bound-state quantum electrodynamics. The result of the experiment on $^{28}$Si$^{13+}$ for the $g$-factor of the single bound electron is $g_{exp} = 1.995\,348\,958\,7\,(5)(3)(8)$ \cite{Stur2011}. The numbers in parenthesis correspond respectively to the systematical and statistical uncertainties and the error of the electron mass at that time. The result agrees well with the theoretical value $g_{theo}= 1.995\,348\,958\,0\,(17)$, which includes QED contributions up to the two-loop level of the order of $(Z\alpha)^2$ and $(Z\alpha)^4$ and represents the so far most stringent test of bound-state QED calculations in strong external fields (see Figure~\ref{GFactor_Contrib}).
\begin{figure*}[h!]
\includegraphics[width=0.7\textwidth]{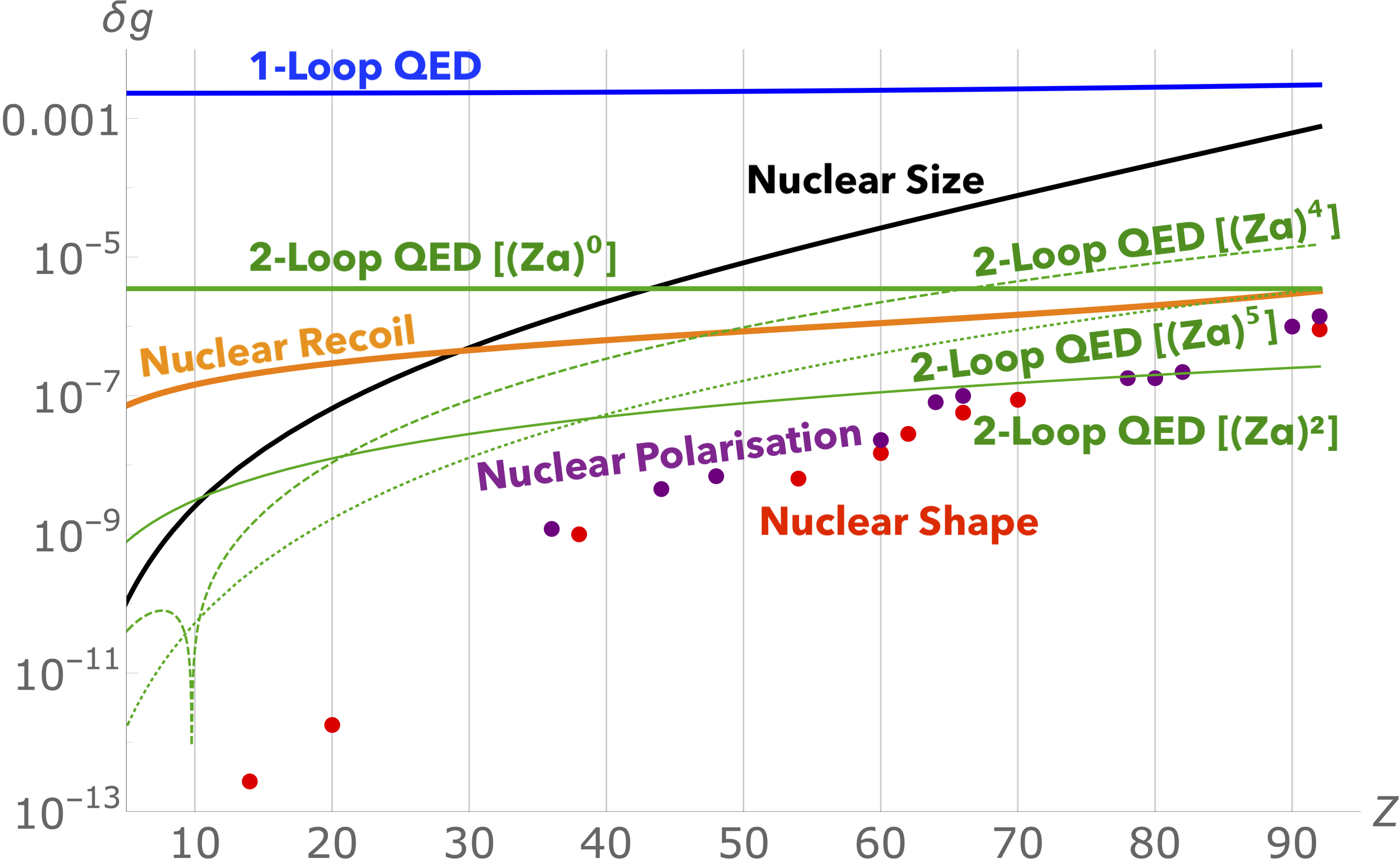}
\centering
\caption{QED (blue and green), nuclear size (black) and recoil (orange) contributions to the $g$-factor bound in hydrogen-like ions for $Z=6$ (carbon) up to $Z=92$ (uranium) \cite{Debierre2020}. The figure is adapted from \cite{PhysRevLett.108.063005} and with results from \cite{PhysRevLett.120.043203,PhysRevLett.89.081802,PhysRevA.99.012505}.}
\label{GFactor_Contrib}
\label{GFactor_Contrib}
\end{figure*}

Similar experiments with comparable precision have been obtained on H-like $^{12}$C$^{5+}$ \cite{Haef2000} and $^{16}$O$^{7+}$ \cite{Verd2004}. Also experiments using lithium like ions $^{40}$Ca$^{17+}$ and $^{48}$Ca$^{17+}$ \cite{Koeh2016}, and $^{28}$Si$^{11+}$ \cite{Wagn2013, Glazov2019} have been performed. Here the interaction of the additional 2 bound electrons modify slightly the value of the $g$-factor. The comparisons between theory and experiment test calculations of the inter-electronic interaction. Most recently the $g$-factor of the boron-like ion $^{40}$Ar$^{13+}$ has been determined with high precision \cite{Arap2019} at the newly commissioned ALPHATRAP experiment at the Max-Planck-Institute for Nuclear Physics in Heidelberg, Germany \cite{Stur2019}. The experimental result distinguishes between disagreeing predictions for the contribution of the electron-electron interaction.

The ALPHATRAP experiment opens the door for ultra-high precision $g$-factor experiments in the highly interesting high-$Z$ regime, and would allow the testing of the most stringently bound-state QED predictions and the achievement of a high sensitivity to physics beyond the Standard Model. The nuclide $^{208}$Pb seems to be the candidate of choice here since it is a heavy ion ($Z = 82$), its nuclear structure is best understood due to its doubly magic character (closed neutron and proton shell), and the high charge state of up to 81+ of these ions provides an about 3 orders of magnitude stronger field compared to silicon, the so far heaviest hydrogenlike system studied in a closed Penning trap. 

\subsection{Measurement of fundamental constants}
The ability to measure the magnetic moment in a large variety of elements and charge states on the other hand allows selecting systems optimised for specific goals. Since the calculation of the $g$-factor, or specifically the magnetic moment of the ion depends on a number of fundamental constants it becomes possible to determine values of those constants by measuring the magnetic moment in a suitable ion or combination of ions and comparing it to the theory value. Of course, in order to do so, it is necessary to trust in the validity of the QED calculation. This, however, is feasible e.g. in the low-$Z$ regime, as we have already confirmed QED in the stronger fields of the medium-$Z$ regime. Furthermore, in the low-$Z$ regime the higher order (in $(Z\alpha)$) contributions of QED are still relatively small and under control. In the past, we have demonstrated this principle by determining the atomic mass of the electron from a measurement of the magnetic moment of hydrogenlike $^{12}$C$^{5+}$ with an unprecedented precision of about 30~ppt. As the Larmor frequency of the bound electron is (anti-) proportional to the electron mass (via the Bohr magneton), but the cyclotron frequency of the ion depends on the ion mass, a $g$-factor measurement relates the electron mass to the ion mass, which can in the case of $^{12}$C$^{5+}$ easily be related to the atomic mass unit $u$. The measurement at the time had been limited by the measurement precision, dominantly by the magnetic field stability. Improvements in this respect make it feasible to revisit this measurement and improve the literature value, which is especially appealing in the light of currently ongoing efforts to achieve a comparable precision with laser spectroscopy of molecular HD$^+$ ions \cite{alighanbari2020precise,patra2017proton}. 
The developed techniques can of course also be applied to direct cyclotron frequency mass spectrometry as discussed in chapter \ref{ch:MassSpec}. For fundamental atomic physics applications the lightest ions, such as the proton, deuteron, tritium and $^3$He as well as the neutron are of special interest. Measurements of these ions however suffer from the large energy-dependent systematic shifts originating from their small mass. This requires a specially optimised, extremely harmonic trap, which also allows detecting the small signals induced by these ions and exquisite control on the radiofrequency excitations and the resulting radii. At the Johannes Gutenberg-University in Mainz we have set up such a system, LIONTRAP \cite{heisse2019high} , which has been recently able to perform previously unprecendented measurements of these light ions  and to unveil important discrepancies in the tabulated literature values \cite{heisse2017high}.     
Another interesting opportunity is a determination of the electromagnetic finestructure constant $\alpha_\textrm{em}$, denoted here simply as $\alpha$. This quantity directly enters the value of $g$ for bound electrons, on the one hand via the QED series solution, but also already on the relativistic level. Since the electron is bound to the nucleus it has to satisfy the Dirac equation for the hydrogenlike atom. The value for $g$ determined from that solution was first calculated by Breit \cite{breit1928magnetic} and depends on $\alpha_\textrm{em}$. Especially for high-$Z$ systems the influence is sizeable and overcomes the impact of the QED contributions. Correspondingly, a determination of $\alpha$ from the bound electron $g$-factor is mostly independent of alternative determinations from the free electron $g$-2 \cite{hanneke2008new} or recoil spectroscopy of atoms \cite{parker2018measurement}, allowing for stringent tests of QED across large variety of physical systems. Especially the $g$-factor of high-$Z$ systems is strongly influenced by the nuclear structure, particularly the nuclear charge radius. On the one hand, this can be exploited in order to determine values for the charge radii with competitive precision as we have shown in \cite{sturm2011g,sturm2013g}, on the other hand the uncertainty of the tabulated nuclear properties poses a limitation on the achievable precision of an $\alpha$ determination in the high-$Z$ regime.
One proposed solution \cite{shabaev2006g} is to use specifically weighted differences of $g$-factors of HCI in different charge states, typically hydrogen- and boronlike. As a 2p-state the boronlike ions have very little overlap with the nucleus, while the 1s state has significantly more. By taking a weighted difference it is thus possible to cancel the small nuclear influence on the boronlike ion using a small admixture of the hydrogenlike $g$-factor. Currently, this procedure is hindered by the challenging theoretical calculation of the 5 electron system to the required precision, but work in this direction is currently actively ongoing. An alternative approach appears when considering low-$Z$ ions. Here, the theoretical precision is markedly higher and the influence of the nuclear properties is considerably smaller, however, also the dependence on $\alpha$ is reduced to an extend that currently the experimentally achievable precision would be insufficient for a competitive determination of  $\alpha_\textrm{em}$. However, using advanced techniques that are currently under development it might be possible in the foreseeable future to determine differences of $g$-factors of two ions with significantly higher precision, rendering the low-$Z$ regime highly interesting.           

\subsection{Measurement of specific $g$-factor differences}
\begin{figure*}[h!]
\includegraphics[width=1\textwidth]{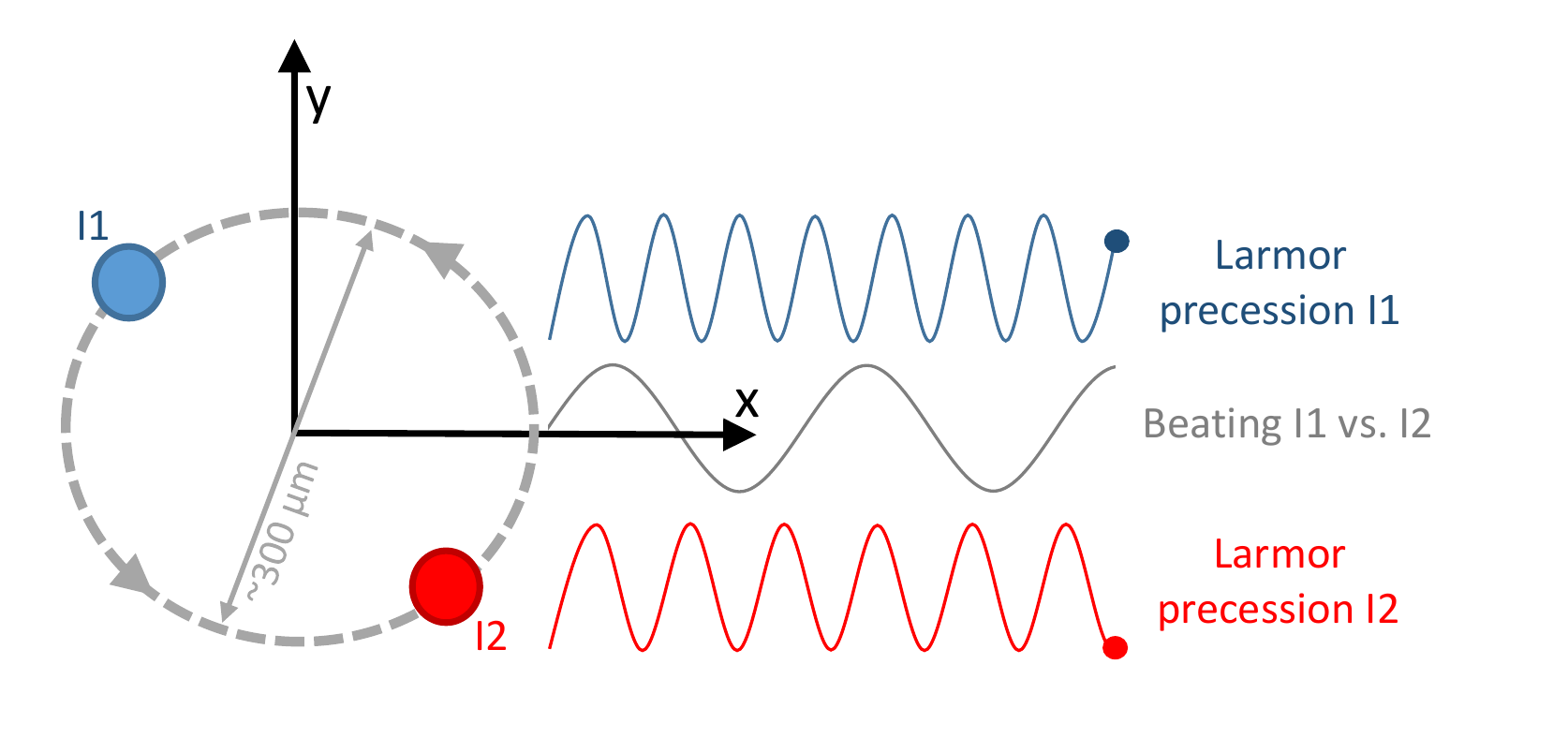}
\centering
\caption{When two ions are stored on a common magnetron orbit they perform a deterministic motion and experience on average almost exactly the same magnetic field. This way, when observing the Larmor precession (using a Ramsey-type interrogation), while the individual (fast) Larmor precessions of the two ions rapidly become incoherent due to magnetic field fluctuations, their beating stays coherent for significantly longer times, provided the two Larmor frequencies are similar. This allows to directly extract the difference of the two similar $g$-factors $\Delta g=g_1-g_2$ with drastically improved resolution compared to individual measurements. }
\label{dgbeating}
\end{figure*}

Generally, as the leading order contributions to the bound electron $g$-factor have been tested already at least in medium-strong fields, it becomes more interesting to investigate also higher-order contributions. Especially in the low-$Z$ regime these are however typically significantly smaller, either due to a strong reduction by factors of  $\bigl( \frac{\alpha}{\pi}\bigr)^n$ and $(Z\alpha)^n$, or by their dependence on nuclear properties. To specifically investigate those contributions, it is advantageous to look at differences of similar systems, where most of the contributions drop out, while the contribution of interest remains. In addition to the above-mentioned determination of $\alpha$, an obvious case is the isotopic effect on the $g$-factor. This is a small contribution (as it depends on the nucleus), but can be observed very clearly by looking at the difference of the $g$-factors of two isotopes in the same charge state. Here, virtually all contributions of the (almost identical) electron shell drop out, while the nuclear contributions, which are typically in the $10^{-8}$ range for medium heavy systems, become clearly visible and enable a unique way to probe the validity of strong field QED beyond the Furry picture, where the nucleus is considered to be immotile \cite{Koeh2016}.    
On the experimental side it is very unfavourable to measure the $g$-factors of both isotopes independently and then to take the difference of the two values, as this requires high precision for both the measurement and the masses of the ions. This precision for the $g$-factor is technically limited by the resolution of the magnetic field achievable by the cyclotron frequency measurement. 
A drastically better result can be achieved if it is possible to use the (much higher) Larmor frequency of a second ion as co-magnetometer. Normally, unavoidable fast magnetic field fluctuations lead to decoherence of the individual Larmor precessions on a time scale of milliseconds. If the two ions however see exactly the same magnetic field, the difference of the Larmor frequency, which is proportional to the sought-after difference of the $g$-factors, stays coherent for much longer times. This is possible by keeping both ions simultaneously in one trap, either in an axial Coulomb crystal, where both ions are (laser-) cooled to stable equilibrium positions along the trap axis, or on a common magnetron orbit, where both ions circulate around the trap axis (see figure \ref{dgbeating}). 
The techniques required to first combine the ions on a common orbit as well as to separate them after the measurement are currently developed. Once the viability of this technique has been shown, a leap in the achievable precision for small contributions to $g$ can be expected, which will consequently lead to intriguing new measurement opportunities. 

\section{Precision laser spectroscopy on highly charged ions}
\begin{figure*}[h!]
\includegraphics[width=1\textwidth]{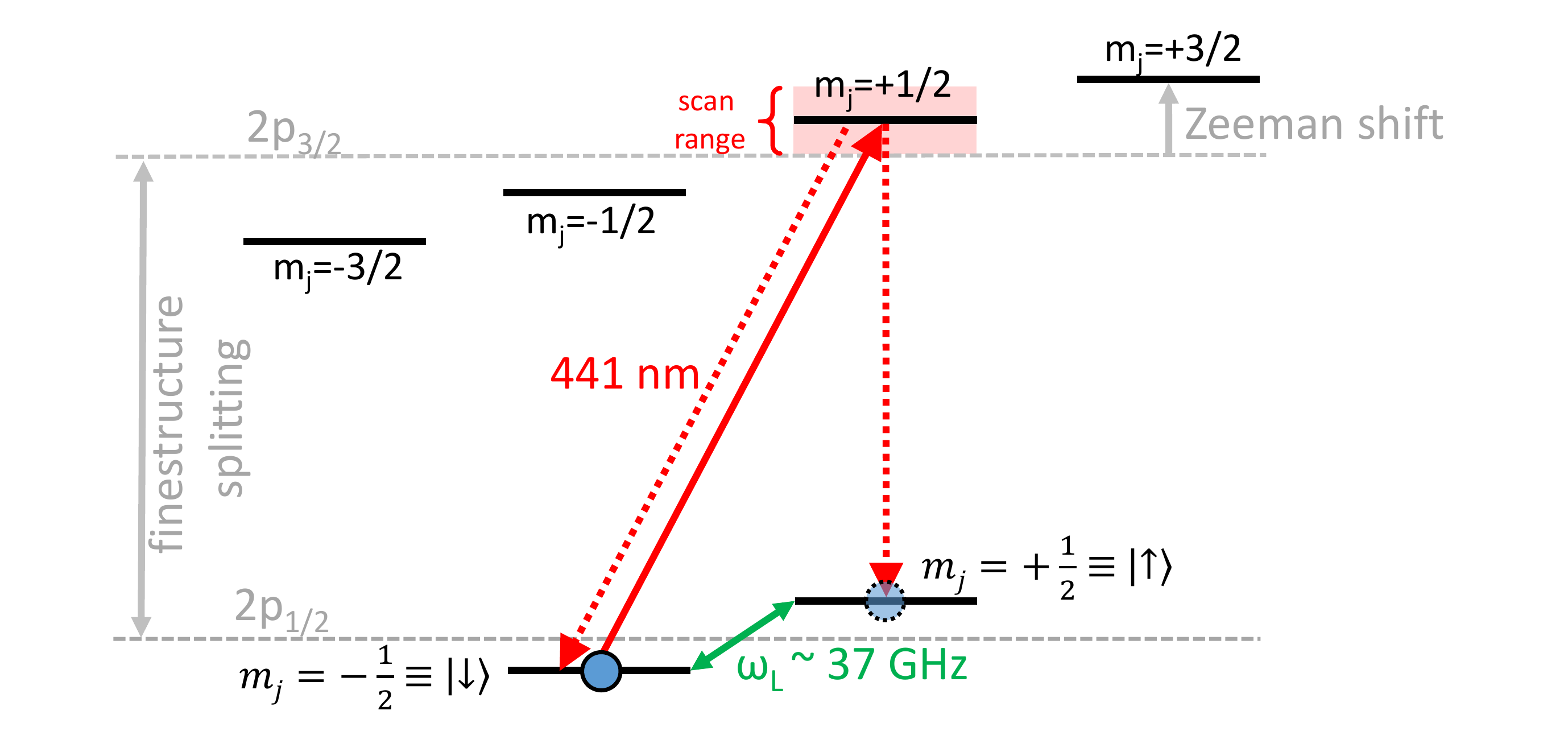}
\centering
\caption{The finestructure of the ground state of boronlike $^{40}$Ar$^{13+}$. By using the continuous Stern-Gerlach effect and millimeter wave excitations at the ground-state Larmor frequency (green arrow), it is possible to prepare the ion deterministically in e.g. the $\left|\downarrow\right>$ state. Then, a laser close to the transition frequency to the 2p$_{3/2}$(m$_j$=+1/2) state can lead (with about 66\% probability) to a decay to the $\left|\uparrow\right>$ state, which afterwards is virtually a dark state due to the detuning by the strong Zeeman shift. The successful transition can then be detected by determining the magnetic substate (now $\left|\uparrow\right>$) with the CSGE. This method of detecting the result of the transition rather than the transition itself thus allows a rapid search for an unknown transition frequency by chirping a large range of frequencies (as indicated in the figure).}
\label{Ar13structure}
\end{figure*}

Many of the most interesting transitions for fundamental atomic physics are at least electric dipole forbidden, as their long lifetimes enable reaching highest precision. However, this long lifetime also goes along with a very low fluorescent yield, rendering standard laser spectroscopy in many interesting cases impossible. Here, the continuous Stern-Gerlach effect (CSGE) offers a unique and versatile way  to  determine and prepare the quantum state of an ion, independent of the lifetime of the respective transition. To this end, the current sub-state of the ion is determined by driving a magnetic Zeeman transition in the ground state. A successful transition can be clearly seen via a change of the axial frequency of the ion, without the need to observe the actual transition. The change of the axial frequency, along with the knowledge of the millimeter-wave frequency used to drive the transition, allows unambiguously identifying the current sub-state. Consequently, if a probe excitation using a laser or any other radiation source successfully drives a transition this can be detected by re-determining the current state (see figure \ref{Ar13structure} for an example). 

A first proof-of-principle of this technique has recently been performed by measuring the finestructure transition in boronlike $^{40}$Ar$^{13+}$, an E1-forbidden transition with about 10~ms lifetime.  Here, the ground-state is a 2p$_{1/2}$ doublet and a successful transition, probed by a 441~nm laser, effectively leads to pumping into the other magnetic sublevel of the doublet. As this method is sensitive to the result (the state change) of the transition rather than the transition itself, it is extremely powerful in locating unknown transition frequencies of narrow transitions. To this end, after preparing the ion in a known sub-state, the laser is chirped over a certain range of frequencies. Afterwards, a change of the sub-state indicates a successful transition within the chirp range, allowing to effectively narrow the search range until the transition is located. 
This new method, combined with the extraordinarily long storage times enabled in the cryogenic vacuum and ultra-low temperatures resulting from sympathetic cooling of the ions of interest will allow in the future to study many systems that have been previously inaccessible to precision laser spectroscopy, such as the hyperfine structure in heavy HCI, e.g. $^{209}$Bi$^{82+}$ \cite{ullmann2017high}, or the ro-vibrational structure of the hydrogen molecular ions, H$_2^+$ and HD$^+$ \cite{alighanbari2020precise,parker2018measurement}, or even its antimatter counterpart $\bar{\textrm{H}}_2^-$ \cite{myers2018c}, once this can be produced in antimatter facilities.

\section{Acknowledgments}
This project was funded by the European Research Council
(ERC) under the European Union’s Horizon 2020 research and
innovation programme under Grant Agreement No. 832848-FunI.
Furthermore, we acknowledge funding and support by the Max-
Planck Society, and funding from the German
Research Foundation (DFG) Research UNIT FOR 
2202 under Project No. DU1334/1-2 and from the Collaborative 
Research Centre “SFB 1225 (ISOQUANT).”

\printbibliography
\end{document}